\documentclass[prl,twocolumn,showpacs,floatfix]{revtex4}
\usepackage{graphicx}
\usepackage{amsmath}

\begin{document}

\title{Phase diagram of a rapidly-rotating two-component Bose gas}
\author{E. \"O. Karabulut$^{1,4}$, F. Malet$^2$, G. M. Kavoulakis$^3$, 
and S. M. Reimann$^1$}
\affiliation{$^1$Mathematical Physics, LTH, Lund University, 
P.O. Box 118, SE-22100 Lund, Sweden \\
$^2$Department of Theoretical Chemistry and Amsterdam Center for 
Multiscale Modeling, FEW, Vrije Universiteit, De Boelelaan 1083, 
1081HV Amsterdam, The Netherlands \\
$^3$Technological Educational Institute of Crete, P.O. Box 1939,
GR-71004, Heraklion, Greece \\
$^4$Physics Department, Faculty of Science, Selcuk University,
TR-42075, Konya, Turkey}

\date{\today}

\begin{abstract}

We derive analytically the phase diagram of a two-component Bose 
gas confined in an anharmonic potential, which becomes exact and 
universal in the limit of weak interactions and small anharmonicity 
of the trapping potential. The transitions between the different 
phases, which consist of vortex states of single and multiple 
quantization, are all continuous because of the addition of the second 
component.

\end{abstract}


\maketitle

One of the remarkable features of cold atomic gases is 
their high degree of tunability, allowing for a precise and
flexible control over most of the experimental parameters,
and paving the way for many different potential applications
of these systems. On the experimental side, remarkable progress 
has been made by, e.g., the realization of confining potentials 
of various functional forms (see, e.g., \cite{tr, dal}), or by 
the creation of mixtures of different atomic species, see, e.g., 
\cite{KTUrev}.

Two-component rotating Bose-Einstein condensates have been
investigated thoroughly for the case of harmonic confinement,  
\cite{KTUrev}. One of the most interesting phenomena is the 
presence of so-called ``coreless vortices'' that occur when 
only one of the two components carries all the angular momentum 
and forms a vortex state around the second one, which remains 
at rest at the vortex core \cite{KTUrev}. Indeed, the presence 
of the second, non-rotating component gives rise to an effective 
anharmonic potential acting on the rotating component, and 
therefore allows the formation of multiply-quantized vortex 
states \cite{Sara, mix}. These multiply-quantized vortex states 
are not energetically favorable in a single-component system in 
the case of harmonic confinement.

The rotational properties of a single-component Bose-Einstein
condensate in the presence of an anharmonic potential have been 
addressed previously, employing the Gross-Pitaevskii mean-field 
approach, or the method of exact diagonalization \cite{Sandy, 
Emil, KTU, FB, KB, LCS, AD, JKL, JK, KJB, BKR, JY}. In the case of 
harmonic confinement the rotational frequency of the trap $\Omega$ 
is limited by the trap frequency $\omega$ because of the centrifugal 
force: as $\Omega \to \omega$ the system enters a highly-correlated 
regime \cite{NC, SFetter, SRHM}, while for $\Omega > \omega$ the 
system is not bounded. On the other hand, in the case of anharmonic 
confinement the system is bounded for any value of $\Omega$.

In this study we consider a mixture of two Bose-Einstein condensates, 
which are confined in an anharmonic potential \cite{Wu}. We investigate 
the rotational properties of this system as a function of the 
rotational frequency of the trap and the coupling between the atoms. 
As we show, in the limit of weak interactions and small anharmonicity 
of the confining potential one can derive the corresponding phase 
diagram analytically solving a quadratic algebraic equation. Remarkably 
the phase diagram is exact and universal in these limits.

Our paper is organized as follows. We first focus on the case
of zero and sufficiently weak interatomic interactions, where the 
order parameters of the two species are multiply-quantized vortex 
states. The simplicity of these states then allows us to investigate 
their stability as the coupling constant between the atoms increases, 
deriving the phase diagram of the system as a function of the 
interaction strength and of the rotational frequency of the trap. 
We finally analyse and interpret our results physically and compare 
them with those of a single-component system, as the inclusion of a 
second component changes the corresponding phase diagram rather 
drastically.

\textit{Model.}\textemdash
We consider a mixture of two distinguishable bosonic atoms,
labelled $A$ and $B$ with equal mass $M$, but with different 
number of atoms $N_A$ and $N_B$. The system is confined in a 
two-dimensional anharmonic potential of the form
\begin{equation}
   V(\rho) = \frac 1 2 M \omega^2 \rho^2 [1 + \lambda (\frac {\rho}
   {a_0})^2],
\label{anh}
\end{equation}
where $\rho$ is the radial coordinate in cylindrical coordinates,
$\omega$ is the trap frequency, $a_0 = \sqrt{\hbar/(M \omega)}$
is the oscillator length, and $\lambda$ is a positive dimensionless
parameter measuring the strength of the anharmonicity of the trapping
potential. Along the axial direction, the density is assumed to be 
homogeneous within a width $Z$, with a total density per unit length 
$\sigma = (N_A + N_B)/Z$. The intra- and inter-species interaction is 
modelled as a hard-core potential with scattering lengths for elastic 
atom-atom collisions $a_{AA}$, $a_{BB}$, and $a_{AB}$, which are 
assumed to be repulsive. The general formalism is given for any value
of the scattering lengths, while the final results are presented
for equal scattering lengths, $a_{AA} = a_{BB} = a_{AB} = a$. The 
dimensionless parameter $\sigma a$ thus gives the ``strength" of 
the interatomic coupling.

\begin{figure}[t]
\includegraphics[width=1\columnwidth]{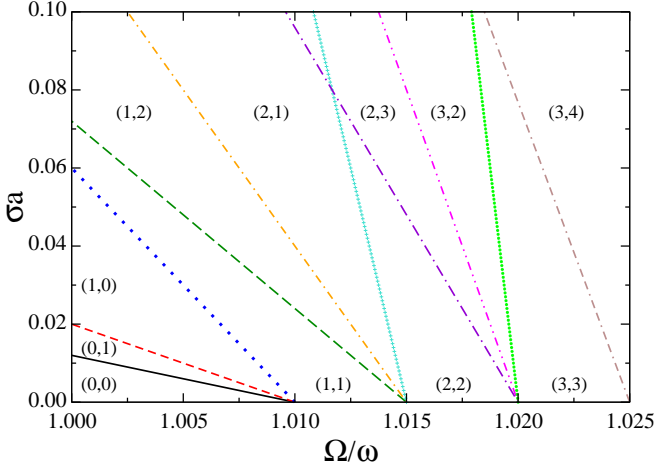}
\caption{(Colour online)
Phase diagram, where the $x$ axis is the frequency of rotation of
the trap $\Omega/\omega$ and the $y$ axis is the coupling $\sigma
a$, for equal scattering lengths. The lines show the discontinuous
transitions between the states $(m,n) \equiv (\Psi_A = \Phi_m; \Psi_B
= \Phi_n$). Here $\lambda = 0.005$ and $N_A/N_B = 2$.}
\label{fig1}
\end{figure}

Within the mean-field approximation the two order parameters $\Psi_A$ 
and $\Psi_B$ obey the following coupled nonlinear, 
Gross-Pitaevskii-like, differential equations, which in the rotating 
frame have the form 
\begin{eqnarray}
  - \frac {\hbar^2 \nabla^2} {2 M} \Psi_i + V(\rho) \Psi_i +
  (g_{ii}|\Psi_i|^2 + g_{ij}|\Psi_j|^2) \Psi_i
\nonumber \\
  - \Omega {\hat L_z} \Psi_i =  \mu_i \Psi_i,
\label{cgpe}
\end{eqnarray}
with $(i = A, B; j = B, A)$. Here ${\hat L_z}$ is the axial 
component of the angular momentum operator, $\mu_{i}$ is the chemical 
potential of each component and $g_{ij} = 4 \pi \hbar^2 \sigma 
a_{ij}/M$. 

\textit{Discontinuous transitions.}\textemdash
Starting with the case of zero coupling, $\sigma a = 0$, and small 
anharmonicity, $\lambda \ll 1$, the order parameters $\Psi_A$ 
and $\Psi_B$ are given by the eigenstates of the harmonic potential 
with no radial nodes $\Phi_m({\rho, \phi}) \propto {\rho}^{m} 
e^{i m \phi} e^{-\rho^{2}/2 a_0^2}$, where $m$ is the quantum 
number that corresponds to the angular momentum $m \hbar$ (assumed
to be positive). The single-particle energy spectrum $E_m$ of the 
above states $\Phi_m$ scales quadratically with $m$, $E_m = \hbar 
\omega [1 + m + \lambda (m+1) (m+2)/2]$, as opposed to the harmonic 
potential, where $E_m \propto m$. As a result, as $\Omega$ increases, 
the system undergoes discontinuous phase transitions between the 
states $\Phi_{m}$. 

For weak coupling, $\sigma a \ll 1$, the effect of interactions may
be treated perturbatively. The energy of the system in a state of 
$(m,n)$, where component $A$ is in the state $\Phi_{m}$, and 
component $B$ is in the state $\Phi_{n}$, is
\begin{eqnarray}
   {\cal E}_{m,n} = x_A E_{m} + x_B E_{n}
\nonumber \\
+ \hbar \omega \sigma a
\left(\alpha_{AA} x_A^2 \frac {(2 m)!} {2^{2 m} (m!)^2}
+ \alpha_{BB} x_B^2 \frac {(2 n)!} {2^{2 n} (n!)^2} \right.
\nonumber \\ \left.
+ 2 \alpha_{AB} x_A x_B \frac {(m + n)!} {2^{m + n} (m! \, n!)}
\right).
\label{intexppp}
\end{eqnarray}
Here, $x_{A,B} = N_{A,B}/N$ and $\alpha_{ij} = a_{ij}/a$. The 
critical frequencies for transitions between different states
of $(m,n)$ can be calculated by comparing the energies in the
rotating frame. Figure 1 shows the corresponding phase boundaries
for a fixed population imbalance $N_A/N_B = 2$ and a weak 
anharmonicity, with $\lambda = 0.005$. In the absence of interactions 
the corresponding critical frequencies are degenerate. However, 
these degeneracies are lifted by the interactions, as seen in 
Fig.\,1. For a fixed coupling and increasing $\Omega$ the system 
first undergoes a (discontinuous) transition to the state where the 
component with smaller number of atoms carries all the angular 
momentum and the other component remains static, which is a so-called 
coreless vortex state. Then, with increasing $\Omega$, a vortex state 
forms in the larger component, while the smaller one becomes static. 
Finally, a vortex state forms in both components for a sufficiently 
large value of $\Omega$. It is interesting that the same result is 
obtained in the case of mixtures in a purely harmonic potential 
\cite{Sara, mix}.

\textit{Continuous transitions.}\textemdash
As the interaction strength increases, the states of multiple 
quantization become unstable, as the energy of the system is 
minimized by mixing states of different angular momentum in the
order parameters $\Psi_A$ and $\Psi_B$. The multiply-quantized 
vortex states undergo continuous, second-order phase transitions. 
It turns out that the order parameters above the phase boundaries 
are of the form
\begin{eqnarray}
 \Psi_A = c_{m} \Phi_{m} + c_{m+q} \Phi_{m+q}, \,
 \Psi_B = d_{n} \Phi_{n} + d_{n+q} \Phi_{n+q}.
\label{inst}
\end{eqnarray}
Sufficiently close to the phase boundary, the coefficients $c_{m}$ 
and $d_{n}$ are of order unity, while the other two coefficients in 
Eq.\,(\ref{inst}) tend to zero. For this reason we keep in the
energy only the terms which are up to quadratic in $c_{m + q}$ and 
$d_{n + q}$,
\begin{eqnarray}
  {\cal E} = \frac {x_A} {S_A} (E_{m} c_{m}^2 + E_{m+q} c_{m+q}^2)+
  \frac {x_B} {S_B} (E_{n} d_{n}^2 + E_{n+q} d_{n+q}^2)
  \nonumber \\
  + \frac {x_A^2} {S_A^2} (c_{m}^4 V_{m,m,m,m}^{AA}
   + 4 c_{m}^2 c_{m+q}^2 V_{m,m+q,m,m+q}^{AA})
   \nonumber \\
  + \frac {x_B^2} {S_B^2} (d_{n}^4 V_{n,n,n,n}^{BB}
   + 4 d_{n}^2 d_{n+q}^2 V_{n,n+q,n,n+q}^{BB})
   \nonumber \\
   + 2 \frac {x_A x_B} {S_A S_B}
   (c_{m}^2 d_{n}^2 V_{m,n,m,n}^{AB}
   + c_{m+q}^2 d_{n}^2 V_{m+q,n,m+q,n}^{AB}
\nonumber \\
   + d_{n+q}^2 c_{m}^2 V_{m,n+q,m,n+q}^{AB}
   + 2 c_{m+q} d_{n+q} c_{m} d_{n} V_{n,m+q,n+q,m}^{AB}),
   \nonumber \\
\end{eqnarray}
where $S_{A} = c_m^2 + c_{m+q}^2$ and $S_B = d_n^2 + d_{n+q}^2$.
Since the ``pure" states $\Psi_A = \Phi_m$ and $\Psi_B = \Phi_n$ 
provide an extremum of the energy at the phase boundaries, the 
first-order derivatives of the energy in the rotating frame, 
${\cal E}_{\rm rot} = {\cal E} - (m+n) \hbar \Omega$, with respect 
to $c_{m + q}$ and $d_{n + q}$ vanish at the phase boundary. 
Therefore, the stability of the states of multiple quantization is 
determined by the eigenvalues of a two by two matrix whose elements 
consist of the second-order derivatives of the energy with respect 
to $c_{m + q}$ and $d_{n + q}$, i.e., $M_{1,1} = \partial^2 {\cal 
E}_{\rm rot}/ \partial c_{m+q}^2$, $M_{2,2} = \partial^2 {\cal E}_{\rm 
rot}/ \partial d_{n+q}^2$, and $M_{1,2} = \partial^2 {\cal E}_{\rm 
rot}/ \partial c_{m+q} \partial d_{n+q}$. The resulting matrix 
elements along the diagonal are 
\begin{eqnarray}
M_{1,1} = 2 x_A (E_{m + q} - E_{m} - q \hbar \Omega)
+ 4 x_A^2 [2 V_{m, m + q, m, m + q}^{AA}
\nonumber \\ - V_{m,m,m,m}^{AA}]
+ 4 x_A x_B [V_{m+q,n,m+q,n}^{AB} - V_{m,n,m,n}^{AB}],
\end{eqnarray}
\begin{eqnarray}
M_{2,2} = 2 x_B (E_{n + q} - E_{n} - q \hbar \Omega)
+ 4 x_B^2 [2 V_{n, n + q, n, n + q}^{BB}
\nonumber \\ - V_{n,n,n,n}^{BB}]
+ 4 x_A x_B [V_{m,n+q,m,n+q}^{AB} - V_{m,n,m,n}^{AB}],
\end{eqnarray}
while the off-diagonal element is $M_{1,2} = 4 x_A x_B 
V_{m, n+q, m+q, n}^{AB}$, where
\begin{eqnarray}
  V_{k,l,m,n}^{ij} = \hbar \omega \sigma a_{ij}
  \frac {(k+l)!} {2^{k+l} \sqrt{k! l! m! n!}}
\, \delta_{k+l,m+n}.
\end{eqnarray}
When all the eigenvalues of this matrix are positive, the system is 
stable. However, as soon as any of the eigenvalues becomes negative, 
an instability occurs via a second-order and continuous phase
transition. Assuming equal scattering lengths, we derive the 
following general expression for the phase boundaries of the 
``pure" states $\Psi_A = \Phi_{m}$ and $\Psi_B = \Phi_{n}$ for 
$n=m$, which is given by
\begin{eqnarray}
  (\sigma a)_q = \frac 1 2 \frac {E_{m+q} - E_m - q \Omega}
  {V_{m,m,m,m} - V_{m,m+q,m,m+q}}.
\label{phase}
\end{eqnarray}
The parameter $q$ in the above expression can take any value, provided 
that $m+q \ge 0$.

The result of the calculation described above is shown in the phase 
diagram depicted in Fig.\,2, which is the main result of this study. 
The phase boundaries corresponding to these continuous transitions 
(triangular regions) remain in between the straight lines of 
discontinuous phase transitions of absolute energetic stability shown 
in Fig.\,1, which implies that the transitions between the phases of 
multiply quantized vortex states are no longer discontinuous. As long 
as the interaction strength has a non-zero value, they are continuous. 
The resulting phase diagram obtained for the two-component case in this 
respect is different from that of the one-component case for which the 
transitions between phases are found to be both continuous and 
discontinuous \cite{JKL}.

The part of the phase boundary which represents the instability 
towards the mixed states with $q=-1$ are vertical due to the fact 
that the denominator of Eq.\,(\ref{phase}) becomes zero (i.e., 
$V_{m,m,m,m} = V_{m,m-1,m,m-1}$) for this unstable mode. The lines 
corresponding to the unstable mode with $q=1$ have negative slope, 
since
\begin{eqnarray}
  (\sigma a)_{q=1} = \frac {m+1} {V_{m,m,m,m}} [1 - \Omega + \lambda 
  (m+2)].
\end{eqnarray}
Thus, the region of stability takes the shape of a triangle for the 
pure state with $m=1$. The lines with positive slope cutting the other 
triangular regions in Fig.\,2 denote the phase boundaries of the most 
unstable mode for the states with $m \ge 2$. Accordingly, when $m=2$ 
and $m \ge 6$, we find that the most unstable mode corresponds to 
$q=-2$. For the intermediate phases with $m=3,4,$ and $5$ there is  
a transient regime where $q=-3$.

As the rotational frequency increases, we observe that the region 
of stability of the multiply-quantized vortex states extends further, 
up to the points where the lines with positive slope cut the 
triangular regions. For sufficiently large values of $m$, demanding 
that $(\sigma a)_{q=1}=(\sigma a)_{q=-2}$ and ignoring terms of order 
of $1/m$, we find that $\Omega \approx 1 + \lambda (m+1)$. In other 
words, for large $m$, the lines with positive slope cut the ones with 
negative slope right at the top of the triangle.

In the obtained phase diagram, we have determined triple points,
where three phases coexist \cite{KB, JKL}. As it is seen from 
Fig.\,2, the states $(2,2)$, $(3,3)$, and $(4,4)$ have such triple
points, where the two phase boundaries with positive and negative
slopes intersect. For example, in the phase $(2,2)$ around the point
where the two corresponding phase boundaries cut each other, there is
a doubly-quantized vortex state, a doubly-quantized vortex state with
a single vortex around it, and two-singly quantized vortex states.
In the fast-rotating regime, $\Omega > \omega$, the effective potential
due to the confinement and the centrifugal potential has a "Mexican 
hat" form, which leads to a hole in the density of the cloud at the 
trap center. This hole appears first at $(4,4)$, as it is seen in 
Fig.\,3, where we show schematic plots of the density and of the phase 
of the two order parameters $\Psi_A$ and $\Psi_B$ for the regions 
above the phases $(2,2)$, $(3,3)$, $(4,4)$, and $(5,5)$. The 
instability of the phase $(3,3)$ is towards $(\Phi_0, \Phi_3)$, 
which has a non-vanishing density at the center of the cloud. On 
the other hand, the instability of the state $(4,4)$ is towards 
$(\Phi_1, \Phi_4)$, which does vanish at the origin. Therefore, 
beyond the phase $(3,3)$ there is always a node in the density of 
the cloud at the center of the trap. From the plots in Fig.\,3 we
also see that the density minima of the one component coincide with 
the density maxima of the other component. The total density of the
cloud thus remains as close as possible to axial symmetry along these 
phase boundaries.

\begin{figure}[t]
\includegraphics[width=1\columnwidth]{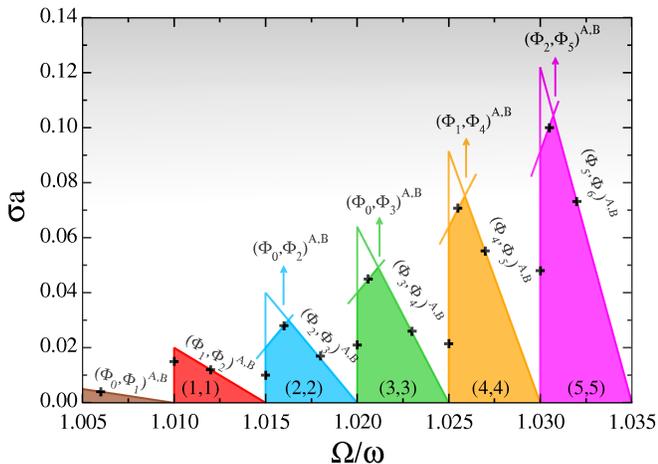}
\caption{(Colour online)
Phase diagram, where the $x$ axis is the frequency of rotation of
the trap $\Omega/\omega$ and the $y$ axis is the coupling $\sigma
a$, for equal scattering lengths. All the lines show continuous
transitions. Here $(m,m) \equiv (\Psi_A = \Phi_m; \Psi_B = \Phi_m$),
while $(\Phi_n, \Phi_m)^{A,B} \equiv (\Psi_{A} = c_n \Phi_n + c_m
\Phi_m; \Psi_{B} = d_n \Phi_n + d_m \Phi_m)$ denotes the states of the 
form of Eq.\,(\ref{inst}). Here $\lambda = 0.005$ and $N_A/N_B = 2$.}
\label{fig2}
\end{figure}

\begin{figure}[t]
\centerline{\includegraphics[width=1\columnwidth]{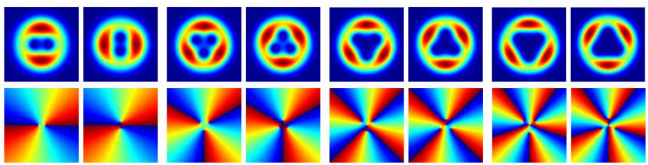}}
\caption{(Color online)
Density (upper) and phase (lower plots) of the component $A$ 
(left) and $B$ (right plots). These plots correspond to the 
``mixed" phases $(\Phi_0, \Phi_2)^{A,B}$, $(\Phi_0, \Phi_3)^{A,B}$, 
$(\Phi_1, \Phi_4)^{A,B}$, and $(\Phi_2, \Phi_5)^{A,B}$ from left
to right.} 
\label{fig3}
\end{figure}

The phase diagram that we have evaluated is exact and universal in 
the limit of weak interactions and weak anharmonicity of the trapping
potential. In the single-component case the phase diagram is also  
universal, however it is only partly exact \cite{JKL} because  
there are also discontinuous phase transitions, where the phase
boundaries can only be evaluated approximately. On the other hand, 
in the present problem all the phase boundaries shown in Fig.\,2 are 
continuous and thus are all exact. The phase diagram is universal in 
the sense that it is invariant under changes of the degree of 
anharmonicity $\lambda$ of the trapping potential (provided that 
$\lambda \ll 1$), under proper rescaling of the two axes.

\textit{Numerical results.}\textemdash
We have confirmed the phase boundaries shown in Fig.\,2 numerically 
for various values of the coupling and of the rotational frequency. 
Since the transitions are continuous, we start from the phases of 
multiple quantization found in Fig.\,1 for a fixed $\Omega$ and 
increase the coupling. We choose the initial state $\Psi_A = \Phi_n$, 
$\Psi_B = \Phi_m$ adding some more states with different angular 
momentum with very small amplitude and then propagate those in 
imaginary time using of a fourth-order split-step Fourier method 
\cite{rel}. Below the phase boundaries the amplitudes of the extra 
components decay, while above the boundary one of the small amplitudes 
increases, indicating the formation of a mixed state. The result of 
this calculation is shown as the crosses in the phase diagram of 
Fig.\,2. The numerical results are in good agreement with the exact 
analytical solution. However, the discrepancy between the results of 
the two methods grows rapidly with increasing $\lambda$, since our 
analytical approach is perturbative.

\textit{Summary.}\textemdash
In summary, we have examined the phase diagram of a mixture of two 
Bose-Einstein condensed gases confined in an anharmonic potential
under rotation, as a function of the strength of the coupling constant 
and of the rotational frequency of the trap. We have shown that it is 
possible to derive the corresponding phase diagram analytically, 
reducing the problem to the evaluation of the roots of an algebraic 
equation of second degree. It is also remarkable that the presence 
of a second component makes the solution of this problem in a sense 
simpler than its one-component counterpart. 

In the case of a single component, for sufficiently weak interactions 
the cloud undergoes discontinuous phase transitions between phases of 
multiple quantization. Here, while there are still phases of vortex 
states of multiple quantization, the transition between them takes
place via continuous transitions. This becomes possible via vortex 
states which enter the two components successively from infinity, 
moving continuously towards the trap center.

The phase diagram we have evaluated is exact for sufficiently weak 
interactions and for small anharmonicity of the trapping potential. As 
long as these two assumptions are not violated, it is also universal.
In the results of our study we have assumed equal scattering lengths 
for inter- and intra-species collisions. In the more general case, 
i.e., when they are not equal to each other (and/or the masses are 
unequal), the phase boundaries are no longer straight lines. However, 
the problem is still of the same level of difficulty. Given the simple 
and systematic behaviour of the evaluated phase diagram, it would be 
interesting to confirm these results experimentally.

We acknowledge discussions with A. D. Jackson. This work was financed
by the Swedish Research Council and originated from a collaboration
within the ``POLATOM" Research Networking Programme of the European
Science Foundation (ESF). E. \"O. K. is supported by the Turkish
Council of Higher Education (Y\"OK) within the scope of the
``Post-Doctoral Research Scholarship Programme".

\end{document}